\documentclass{article}
\begin{document}
\title{ Propagation of electromagnetic waves in space plasma.}

\author{Jerry Jensen\footnote{ ATK Thiokol Propusion, (Independant research). ; email : Jerry.Jensen@ATK.com}\and Jacques Moret-Bailly 
\footnote{Laboratoire de physique, Universit\'e de Bourgogne, BP 47870, F-21078 Dijon cedex, France. email : Jacques.Moret-Bailly@u-bourgogne.fr}}
\maketitle

\begin{abstract}
  Coherent Raman Effect on Incoherent Light  (CREIL), shifts the frequencies of normally incoherent light without any blurring of the images or altering the order of the spectra. CREIL operates in gases having 
quadrupolar resonances in the megaherz range, and it is easily confused with Doppler effects. When CREIL is taken into account, the propagation of light in cosmic low pressure gases involves a complex 
combination of absorptions and frequency shifts. The propagation of light in the extended photosphere of  extremely hot objects is so complex  because CREIL requires a Lyman excitation in atomic hydrogen to 
achieve hyperfine resonance states in the first excited quantum levels of atomic hydrogen. A bistability emerges which chains Lyman absorption patterns through a coincidence of  lines from each pattern allowed at 
one redshift  into line patterns that coincide at other redshifts  This is not a coincidence of whole patterns, only of one line (for instance the coincidence of a redshifted Ly beta on the Ly alpha of the gas).  This 
chaining is reflected in the apparent harmonic distribution of quasars.

Current star theory predicts very bright accreting neutron stars. These should be small, very hot objects surrounded by dirty atomic hydrogen.  CREIL predicts spectra for these stars that have exactly the 
characteristics found in the spectra of the quasars. The intrinsic redshifting in the extended photosphere of Quasars as defined by CREIL events drastically reduces both the size and distance to quasars, and clearly 
identifies the missing neutron stars as quasar-like objects. A full interpretation of quasar spectra does not require jets, dark matter,  a variation of the fine structure constant, or an early synthesis of iron. CREIL is 
useful in explaining other astrophysical problems, such as redshifting proportional to the path of light through the corona of the Sun. CREIL radiation transfers  may explain the blueshifting of radio signals from 
Pioneer 10 and 11.

\end{abstract}

\section{Introduction. }

Coherent light-matter interactions are usually associated with high Planck temperatures, such as in lasers operation. However, these interactions occur often at 
all intensities of light. For example, a change in the density of the media, a light pulse is propagating in, results in refraction, a coherently scattered Rayleigh 
mode which occurs at even the lowest detectable intensities.

Incoherent Rayleigh scattering (such as the blue of the sky) is generally coincidental with incoherent Raman scattering. There are naturally occurring conditions 
where Raman scattering is coherent. This is "Coherent Raman Effect on Incoherent Light" (CREIL). The first part of section \ref{recall} summarizes the theory of 
CREIL, more precisely developed in previous papers  \cite{Mor98a,Mor98b,Mor01}. The second part of this section compares the properties of the CREIL with 
the properties of its avatar,  Impulsive Stimulated Raman Scattering  (ISRS) \cite{Giordmaine,Yan,Weiner,Dougherty}. ISRS is easily obtained in the laboratory. 
Section \ref{homog} develops the result of a previous paper \cite{Mor03} indicating that the CREIL introduces a bistability during the propagation of the high flux 
of light radiated by an extremely hot source into low pressure atomic hydrogen. Section \ref{spectrum} shows that this bistability produces a multiplication of 
absorption lines (The Lyman forest).

Sections \ref{halo} and  \ref{pion} demonstrate applications of the CREIL in astrophysics: Section  \ref{pion} applies the theoretical results obtained in section 
\ref{spectrum} to a cloud of dirty hydrogen surrounding an accreting neutron star. The very complicated spectrum which is obtained has the properties found in 
quasar spectra, solving both the question of the origin and nature of the quasars; and the lack of observation of accreting neutron stars, (which should be easily 
detectable, but have not been identified). Section 6 is an application of the CREIL to the solar plasma, explaining the broadening of lines in the corona (Wilson-
Bappu effect \cite{Cardini}), and the anomolous blueshift of the radio transmissions of the Pioneer 10 and 11 space probes.

\section{A Summary of the "Coherent Raman Effect on  Incoherent Light" (CREIL).}\label{recall}

A difference in phase exists between an exciting monochromatic beam, and the same beam scattered by molecules without a change of frequency (Rayleigh 
scattering). Rayleigh scattering is not time dependant, so that both the scattered and the incidental beams remain coherent after collisions, and interfere with 
each other to produce refraction. These collisions introduce only small perturbations. This is the source of Rayleigh incoherent scattering observed in sunlight.

If the scattered light has a different frequency, the difference in phase increases linearly with the time. Since each collision restarts the scatterings, they reset 
the initial phase, introducing an incoherence between the light scattered by different molecules. To prevent this incoherence, a coherent effect requires light 
pulses that are shorter than the collisional moment (the mean time between two collisions). 

The interference of two laser beams (or of the beams of a Michelson interferometer with a moving mirror) produces beat frequencies. If the duration of the 
observation is short, or if short pulses are used, the beats are not detectable and a spectrometer then sees a single, intermediate alias frequency. An 
elementary computation shows that the obtained frequency is in proportion to the amplitudes of the mixed beams. It is worth noting Doppler effects due to the 
thermal or turbulent speed of the molecules is negligible, corresponding to the Raman frequency. Under both conditions, the light pulses are shorter than the 
collisional moment, and shorter than the period of the beats (the period of the Raman resonance), define \textquotedblleft ultrashort pulses\textquotedblright as 
defined by G. L. Lamb as \textquotedblleft shorter than all relevant relaxation times\textquotedblright \cite{Lamb}.

The above explanation is simplistic and flawed. This is very evident when the intensity of the light is low: When light propagates into a medium, the medium is 
polarized (allowing a non-destructive observation of the light). This means, that light increases the energy of the medium by a small displacement of the energy 
levels, rather than a transition. In refraction, this energy is provided by the pulse of light while its amplitude increases, and returned when the amplitude 
decreases. This is not possible with a simple Raman scattering, therefore there must be second radiative process which allows a return of the molecule to its 
stationary state. The simplest process is an opposing Raman scattering.  It is from these concepts: Coherent Raman scattering and the tensors of polarization, 
that the rigorous definition of the Coherent Raman Effects on Incoherent Light (CREIL) is obtained:

\medskip
{\it CREIL is a coherent interaction between a set of pulsed electromagnetic modes which increases their entropy by a transfer of energy from the 'hot' modes to 
the 'cold' modes, and a corresponding red (resp. blue) shift of the hot (resp. cold) modes. This interaction requires as a 'catalyst', a low pressure gas, in 
which the light transition pulses are shorter than the kinetic collisional rate of the gas, and also shorter than the periods of quadrupolar resonances. Since this is 
a coherent effect, CREIL does not blur the images.} 

\medskip
Very often the beam temperatures are very different, so that CREIL is generally limited by the optical properties of the gas. Assuming that the dispersion of 
the polarisability is neglected, the relative frequency shifts $\Delta\nu/\nu$ do not depend on the frequencies $\nu$, so that CREIL frequency shifts may be 
easily confused with Doppler frequency shifts.

The simplest way to meet the conditions necessary for CREIL is by using ultrashort laser pulses. With femtosecond lasers, this experiment is easy to 
demonstrate, but the peak intensity needed to detect the light is so large that the frequency shift is also a function of this intensity: This is known as  Impulsive 
Stimulated Raman Scattering   (ISRS), in which the frequency shift is nearly proportional to the intensity.
 
ISRS may be observed in the laboratory using condensed matter and infrared frequencies while CREIL requires low pressure gases and hyperfine (or 
equivalent) frequencies. Thus, the difference of the lengths of the pulses between lasers (femtoseconds) and ordinary incoherent light (of the order of 5 
nanoseconds) drastically increases the path length necessary to observe this phenomenon, rendering experimental conformation of this process in the 
laboratory impractical. However, a space-based demonstration has already been unwittingly conducted (see section 6).

\section{ Propagation of Bolometric Light in hot, homogenous hydrogen. }\label{homog}
To create a CREIL effect in an extended photosphere, it is first assumed the gas pressure has been reduced to where the rms time between kinetic collisions is 
longer than the average length of the pulses, of the order of 5 nanoseconds. The thermal radiation provides the cold transfer mode.

\medskip
 In a very hot zone, the atoms are ionised; there are no absorption or emission by the protons, but heavier atoms emit spectra.

As the medium cools,  the ionization and the thermal excitation of the atoms deminishes. The hydrogen Lyman emission or absorptions lines begin to form. In 
their ground quantum state (principal quantum number $n = 1$), the neutral atoms have the well known 1420 MHz resonance, the excitation period of which is 
too short to allow a CREIL. But, the resonances corresponding to the quadrupole allowed transitions ($\Delta$F = 1) that have the following frequencies 
178MHz in the 2s$_{1/2}$ spin state, 59MHz in 2p$_{1/2}$ spin state, and 24MHz in 2p$_{3/2}$ are low enough to allow CREIL, and high enough to produce a 
strong CREIL effect. With each increase in principal quantum number, the Lyman absorption and the quadrupolar resonance frequencies decrease rapidly, so 
the CREIL effects are usually only active in the second and third quantum numbers.

As a first approximation of quantum effects, assume only a single CREIL mechanism after a Lyman $\alpha$ absorption. Now assume that the propagation of 
the light along a path $L$ produces a frequency shift $\Delta\nu$ of the Lyman line, the displacement of the shift being much greater than the width of the line. 
This shift occurs only when the atoms are pumped by Lyman absorptions, to low values of the final quantum number n. The resulting redshift is proportional to 
the number (per unit of area) of excited atoms. Assuming that the decay time of the excited states is constant, the energy absorbed over $\Delta\nu$ has a well 
defined value that is proportional to $\Delta\nu$. The subtraction of the corresponding constant, (characteristically constant) intensity $I_c$ from the spectrum 
previously written into the light increases the contrast of this spectrum. On the other hand, the absorptions which occur during the redshift cause wide, weak, 
unobservable lines.

This supposition fails if $I_c$ is greater than the incident intensity, that is if the absorption of the energy in a line width is unable to produce a redshift greater 
than the $\alpha$ line width. In this case, after a short path in the gas, it is no longer possible to achieve Lyman pumping, and there is a termination of this 
redshifting mode. All lines, in particular the lines of impurities in the medium are written into the spectrum by this process. The contribution to CREIL by other 
Lyman absorptions does little to alter the previous description. For instance, it may be necessary to have a full Lyman beta absorption to cause a succinct 
disappearance of the redshift; but this does not extend to higher quantum numbers because the efficiency of the CREIL decreases.

\medskip
{\it To summarize, the bistability of the interaction induces two types of propagation: If the intensity is greater than $I_c$, the light is redshifted so that no 
spectral lines are visibly written into the spectrum. If the intensity is less than $I_c$, there is very little redshifting, all spectral lines of the gas are visibly 
written into the spectrum.}

\medskip
Now assume that the hydrogen is heated by the absorption. If the temperature becomes high enough to pump atoms into excited states, the heating
increases the redshifting power of the gas , allowing a pumping of a higher intensity by previously not absorbed regions of the light spectrum; this induces 
an even greater heating: a new contribution to the bistability appears.

\section{Multiplication of the spectral lines by a propagation in atomic, homogenous hydrogen.}\label{spectrum}
Assume that the intensity in the light spectrum is generally greater than $I_c$. In this case, there is usually a redshift which blurs the absorptions lines. If it so 
happens that an absorption line for which the intensity is lower than $I_c$ is redshifted to superimpose on a strong Lyman absorption line, the redshifted line is 
fully absorbed by the Lyman absorption, the redshift stops, and the whole spectrum of the gas may be written into the spectrum. The coincidence of the 
written, redshifted Lyman $\beta$ (resp. Lyman $\gamma$) line with the Lyman $\alpha$ line of the gas writes the Lyman  pattern into the spectrum of the light. The initial 
and just written Lyman patterns differ by the shift of frequencies $\nu_{(\beta \,{\rm resp.}\, \gamma)}- \nu_\alpha$ between the $\alpha$ and $\beta$ \,(resp. \,$\gamma$) 
lines. As in the standard computations the  lines are considered as Lyman $\alpha$, the frequency shift is relative to the Lyman $\alpha$ frequency:

\begin{equation}
\begin{array}{l}
\qquad z_{(\beta \,{\rm resp. }\,\gamma) , \alpha}=\frac{ \nu_{(\beta \,{\rm resp.} \,\gamma)}-\nu_\alpha}{  \nu_\alpha} \approx \\
\approx\frac{1-1/(3^2 \,{\rm resp. \,}4^2)-(1-1/2^2)}{1-1/2^2}
\end{array}
\end{equation}

\begin{equation}
\begin{array}{l}
z_{(\beta , \alpha)}\approx 5/27 \approx 0.1852 \approx 3*0.0617; \\  z_{(\gamma , \alpha)}= 1/4 = 0.25 =4*0.0625.
\end{array}
\end{equation}

Similar to $ z_{(\gamma , \beta)}\approx 7/108 \approx 0.065.$ The redshifts appear, with a good approximation as the products of $z_b = 0.062$ and an integer $q$, the 
resulting pattern of harmonic intensities is observed in the spectra of the quasars (Burbidge \cite{Burbidge}, Burbidge \& Hewitt  \cite{Hewitt}, Tifft \cite{Tifft76,Tifft95}, 
Bell \cite{Bell}, Bell \& Comeau \cite{Comeau}).

The intensities of the Lyman lines are decreasing functions of the final principal quantum number $n$, so  that the inscription of a pattern is better for $q =3$ than for $q = 4$ 
and {\it a fortiori} for $q = 1$.

\medskip Iterating, the coincidences of the shifted line frequencies with the Lyman $\beta$ or $\alpha$ frequencies  build a  \textquotedblleft tree\textquotedblright , final values 
of $q$ being sums of the basic values 4, 3 and 1. Each step being characterised by  the value of q, a generation of successive lines is characterised by successive values of $q: 
q_1, q_2...$ As  the final redshift is $q_F*z_b = (q_1 +q_2 +...)*z_b $, the addition $q_F = q_1 + q_2 +...$ is both a  symbolic representation of the successive elementary 
processes, and the result of these processes.

 The metaphor \textquotedblleft tree\textquotedblright, is not very good because  \textquotedblleft branches\textquotedblright  of the tree may be  \textquotedblleft 
stacked\textquotedblright  by coincidences of frequencies. A remarkable coincidence happens for $q = 10$, this number  being obtained by the effective coincidences deduced 
from:

\begin{equation} 10=3+3+4=3+4+3=4+3+3=3+3+3+1=... \end{equation} 
$q = 10$ is so remarkable that $z_f = 10z_b = 0.62$ may seem experimentally a value of $z$ more  fundamental than $z_b$.

 \medskip
In these computations, the levels for a value of the principal quantum number $n$ greater than 4 are neglected by reasoning that the corresponding transitions 
are too weak. It is worth noting that an emission line written over the continuous spectrum increases the redshift, so that it decreases the absorptions: the 
coincidence of a written emission line with a Lyman line seems to induce emission lines, so that the previous description applies to the emission lines as well.

\section{Propagation of the light in the halo of a hot star.}\label{halo}

First assume that this halo is made mostly of atomic hydrogen, and we generally do not take the impurities into account. 

Close to the star, the pressure and temperature are high and the atoms of hydrogen are fully ionised, there is no CREIL and no redshifting. Since a star is not a 
blackbody, the gas is hotter than the light (just as in the corona of the Sun), the emission lines of the (ionized) impurities appear sharp. At greater distances, 
the multiplication of the lines described in section 4 comes into play.

At a small distance, the gas is relatively dense and hot enough for some thermal population of the excited states of the hydrogen. Therefore, there is always 
some redshift manifest as line broadening. As the temperature decreases, the emission lines disappear, and after a no-line region, which corresponds to an 
approximate equality of the temperatures of the gas and the light, the Lyman absorption lines appear. The saturation of the lines due to the relatively high 
density of gas corresponds to an equilibrium of temperature between the gas and the light at the top of the broad lines; these lines get the shape of a trough.

This description fails if the star radiates a strong radio field. This accelerates the free electrons, ionizing the gas in these relatively high density regions. This is 
why radio-loud stars do not have broad lines. At a very great distances, after creating the spectral Lyman forest, the Lyman absorption disappears, the hydrogen 
cools and becomes diatomic. All redshifted wavelengths provide energy for blue shifting of low energy frequencies in the thermal and radio bandwidths. These 
converge in the infrared frequency range, making \textquotedblleft very distant\textquotedblright  objects appear very dusty.  

\section{Observations of the CREIL in the solar system.}\label{pion}

The variation of frequency shifts on the solar disk shows three origins for these shifts: Doppler, gravitational, and a shifting proportional to the path of the light 
through
the corona. CREIL effects easily explains the corona line shifting and the resulting Wilson-Bappu effect \cite{Cardini} in stellar spectra. 

Another result of CREIL is evident in a crucial experiment \cite{Anderson}: Fixed Doppler radio signals to or from Pioneer 10 and 11 are blueshifted. Assume 
that the solar plasma between these Pioneer probes and the Earth contains molecules possessing resonances in the megaherz range (either or for instance 
Lyman pumped atomic hydrogen). These molecules transfer by a CREIL process, energy from the solar radiation in the optical range to the thermal radiation 
close to 2.7K. The energy of the radio waves in the optical mode selected by the radio-telescopes is increased by the emitters, over the 2.7K radiation 
energy which makes the background noise, just enough to allow the detection. This leads to a net blue shifting of the signals during radiation transfers.

Crucial experiments could be performed: the blueshift should disappear if a
continuous, constant amplitude wave were emitted. Studying the variations of the blueshift as a function
of the length of emitted pulses, a spectroscopy of the quadrupolar resonances in
the solar plasma could be tried.

\section{Conclusion}

The displacement of quasars by intrinsic CREIL redshifting has prevented the proper assessment of their fundamental nature. We have demonstrated how the 
spectrum of quasars contains the signature of the missing   accreting neutron stars  , (Treves and Colpi  \cite{Treves}, Popov et al \cite{Popov}). 

 A previously proposed new model of the quasar \cite{Mor03} is not conceptually wrong, but more complicated and less precise than this one. Our present 
model uses only the consequences of the standard spectroscopy and matches many phenomonon as observe in nature. This is especially true when compared 
with the standard cosmological model. Specifically:

- Since most of the redshift is intrinsic, quasars are not as distant as current estimates. Consequently, when the Hubble law is applied to the remaining 
redshifted lines \cite{Petitjean,Shull}, they are much closer, less massive, and less brilliant than current theory dictates. 

- The complexity of quasar spectra, including the shapes of the lines, is best explained by CREIL effects in a simple dirty hydrogen halo, as opposed to a 
sequence of baffling jets and sporatic clouds strained by dark matter \cite{Tytler}.

- Radio-loud quasar spectra demonstrate strong absorptions close to the redshift of the emission lines, these lines are intrinsically redshifted into broad lines in 
radio-quiet quasars \cite{Briggs, AndersonS}.

- In a Big Bang scenario, since the neutron stars [quasars] are not older than the local universe, no early generation of iron is needed to explain these spectra.

- Since CREIL depends on the polarisability of an active gas, the dispersion of this polarisability produces variations of the relative frequency shifts 
$\Delta\nu/\nu$ which distort the multiplets: there is no need to vary the fine structure constants \cite{Webb}, to acheive these effects. 

- The most redshifted and bright objects, in particular the broad absorption line (BAL) quasars, emit a relatively hot thermal spectrum  \cite{Omont} synthesized 
by CREIL.

- The Seyfert galaxies may actually be accreting neutron stars blurred by exceptionally thick inhomogeneous clouds.

CREIL is necessary to understand the variations of redshifts in the photosphere of the sun, and the anomalous blueshifts of the radio emissions of the Pioneer 
space probes. It is a powerful tool in understanding the propagation of the light in low pressure gases, and to understanding many other astronomical 
observations.


\begin{thebibliography}{}
\bibitem{Mor98a}Moret-Bailly, J., \textquotedblleft Un effet param\'etrique en astrophysique\textquotedblright, {\it Ann. Phys. Fr.}, {\bf 23}, C1-235-C1-236, 1998.
\bibitem{Mor98b}Moret-Bailly, J., \textquotedblleft Correspondence of classical and quantum irreversibilities\textquotedblright, {\it Quantum and Semiclassical Optics}, {\bf 
10}, L35-L39, 1998.
\bibitem{Mor01}Moret-Bailly, J., \textquotedblleft Influence of the time-coherence of light on the absorption lineshapes of low-pressure gases\textquotedblright, {\it J. Quant. 
Spectr. \& rad. Transfer}, {\bf 68}, 575-582, 2001.
\bibitem{Giordmaine}Giordmaine, J. A., M. A. Duguay \& J. W.Hansen, \textquotedblleft Compression of optical pulses\textquotedblright, {\it IEEE J. 
Quantum Electron. }, {\bf 4}, 252-255, 1968.
\bibitem{Yan}Yan Y.-X., E. B. Gamble  Jr. \& K. A. Nelson, \textquotedblleft Impulsive stimulated scattering: General importance in femtosecond laser pulse interactions with 
matter, and spectroscopic applications\textquotedblright, {\it J. Chem Phys. } {\bf 83}, 5391-5399, 1985.
\bibitem{Weiner}Weiner, A. M., D. E. Leaird., G. P. Wiederrecht \& K. A. Nelson, \textquotedblleft Femtosecond pulse sequences used for optical manipulation of molecular 
motion\textquotedblright, {\it Science}, {\bf 247}, 1317-1319, 1990.
\bibitem{Dougherty}Dougherty, T. P., G. P. Wiederrecht, K. A. Nelson, M. H. Garrett, H. P. Jenssen \& C. Warde, \textquotedblleft Femtosecond resolution of soft mode 
dynamics in structural phase transitions\textquotedblright, {\it Science}, {\bf 258}, 770-774, 1992.
\bibitem{Mor03} Moret-Bailly, J., \textquotedblleft Propagation of light in low pressure ionised and atomic hydrogen. Application to astrophysics. \textquotedblright, {\it 
IEEETPS, special issue on astrophysics}, december 2003.
\bibitem{Cardini}Cardini, D., A. Cassatella,.M. Badiali, A. Altamore, M. J. Figueroa, \textquotedblleft A study of the Mg II 2796.34 A emission line in late--type normal, and 
RS CVn stars\textquotedblright, Astro-ph 0307296, 2003.
\bibitem {Lamb}G. L. Lamb Jr., \textquotedblleft Analytical descriptions of ultrashort optical pulse propagation in a resonant medium\textquotedblright, {\it Rev. Mod. Phys. }, 
{\bf 43}, 99-124, 1971.
\bibitem{Rauch}Rauch, M., W. L. W. Sargent, T. A. Barlow, \textquotedblleft Small-scale structure at high redshift\textquotedblright, {\it ApJ}, {\bf 515}, 500-505, 1999.
\bibitem{Burbidge}Burbidge, G., \textquotedblleft The distribution of redshifts in quasistellar objects, N-systems and some radio and compact galaxies\textquotedblright, {\it 
ApJ.}, {\bf 154}, L41-L45, 1968.
\bibitem{Hewitt}Burbidge, G. \& A. Hewitt, \textquotedblleft The redshift peak at z=0.06. Intrinsic redshifts and the Hubble constant\textquotedblright, {\it ApJ.}, {\bf 359}, 
L33-L36 1990.
\bibitem{Tifft76}Tifft, W. G., \textquotedblleft Discrete states of redshift and galaxy dynamics\textquotedblright, {\it ApJ.}, {\bf 206}, 38-56, 1976.
\bibitem{Tifft95}Tifft, W. G., \textquotedblleft Evidence for quantized and variable redshifts in the cosmic background rest frame\textquotedblright, {\it Astrophys. Space Sci. 
(Netherlands)}, {\bf 244}, 29-56, 1996.
\bibitem{Bell}Bell, M. B., \textquotedblleft Evidence that an intrinsic redshift component that is a harmonic of z = 0.062 may be present in every quasar 
redshift\textquotedblright , arXiv:astro-ph/0208320, 2002.
\bibitem{Comeau}Bell, M. B. \& S. P. Comeau, \textquotedblleft Intrinsic redshifts and the Hubble constant\textquotedblright , arXiv:astro-ph/0305060, 2003.
\bibitem{Anderson}Anderson, J. D., P. A. Laing, E. L. Lau, A. S. Liu, M. M. Nieto \& S. G. Turyshev, \textquotedblleft Study of the anomalous acceleration of Pioneer 10 and 
11\textquotedblright, arXiv:gr-qc/0104064, 2002.
\bibitem{Treves}Treves, A. \& M. Colpi, \textquotedblleft The observability of old isolated neutron stars\textquotedblright, {\it Astron. Astrophys.}, {\bf 241}, 107-111, 1991.
\bibitem{Popov}Popov, S. B., A. Treves \& R. Turolla, \textquotedblleft Radioquiet isolated neutron stars: old and young, nearby and far away, dim and very 
dim\textquotedblright, arXiv:astro-ph/0310416, 2003.
\bibitem{Petitjean}Petitjean, P., R. Riediger \& M. Rauch, , \textquotedblleft The metal line systems inHS1700=6416 : Evidence for inhomogeneities\textquotedblright, {\it 
A\&A}, {\bf 307}, 417-423, 1996.
\bibitem{Shull}Shull J. M., J. T. Stocke \& S. Penton, \textquotedblleft Intergalactic hydrogen clouds at low redshift: Connections to voids and dwarf galaxies\textquotedblright, 
{\it AJ}, {\bf 111}, 72-84, 1996.
\bibitem{Tytler}Tytler, D., \textquotedblleft The distribution of QSO absorption system column densities: evidence for a single population\textquotedblright, {\it ApJ}, {\bf 
321}, 49-79, 1987.
\bibitem{Briggs}Briggs, F. H., D. A. Turnshek \& A. M. Wolfe, , \textquotedblleft The broad absorption lines in spectrum of the QSO PKS 1157-014 : a possible link between 
broad absorption lines QSO's , metal enrichment, and the formation of galaxies\textquotedblright, {\it ApJ}, {\bf 287}, 549-554, 1984.
\bibitem{AndersonS}Anderson, S. A., R. J. Weymann, C. B. Foltz \& F. H. Chaffee Jr., \textquotedblleft Associated CIV absorption in radio-loud QSOs. The '3C mini survey' 
\textquotedblright, {\it AJ}, {\bf 94}, 278-288, 1987.
\bibitem{Webb}Webb J. K., V. V. Flambaum, C. W. Churchill, M. J. Drinkwater \& J. Barrow, \textquotedblleft Millimeter emission from high redshift radioquiet 
quasars\textquotedblright, {\it Phys. Rev. Lett.}, {\bf 82}, 884,1999.
\bibitem{Omont}Omont, A., R. G. McMahon, J. Bergeron, P.Cox, S. Guilloteau, E. Kreysa, F. Pajot, E. Pecontal, P. Petitjean, P. M. Solomon \& L. J. Storrie-Lombardi, 
\textquotedblleft Millimeter emission from high redshift radioquiet quasars. \textquotedblright, {\it Early Universe with VLT. Proc. of the ESO workshop, Garching, Germany, 
1-4 April1996. Springer -verlag}, 357-360, 1997.
\end{thebibliography}
\end{document}